\documentclass[a4paper]{jpconf}
\usepackage{graphicx}
\usepackage{bm}
\begin{document}
\newcommand{\eqref}[1]{(\ref{#1})}
\title{Effective quantum pseudospin-$1/2$ model for Yb pyrochlore oxides}

\author{Shigeki Onoda}

\address{Condensed Matter Theory Laboratory, RIKEN, 2-1, Hirosawa, Wako 351-0198, Saitama, JAPAN}

\ead{s.onoda@riken.jp}

\begin{abstract}
An effective quantum pseudospin-$1/2$ Hamiltonian for Yb$_2TM_2$O$_7$ ($TM$ = Ti and Sn) is obtained in terms of the atomic Kramers ground doublet of the $LS$ coupling and the crystalline electric field, which is almost described with $J^z=\pm1/2$. It is calculated microscopically as the sum of Anderson's superexchange interaction and the magnetic dipole interaction. It is found that it shows a strong exchange anisotropy. 
\end{abstract}

\section{Introduction}

Geometrical frustration and quantum fluctuation provide a promising route to unconventional states in magnetism, including various types of spin liquids characterized by the absence or emergence of unconventional orders of chirality or multipoles in three spatial dimensions~\cite{anderson:56,anderson,wen:89}. 
The pyrochlore lattice structure offers a prototypical stage where the geometrical frustration prevents the magnetic dipole long-range order (LRO)~\cite{anderson:56,moessner:98,isakov:04}. One class of magnetic pyrochlore oxides~\cite{gardner:10} which have been most intensively studied is a dipolar spin ice, e.g., Dy$_2$Ti$_2$O$_7$ and Ho$_2$Ti$_2$O$_7$~\cite{gardner:10,harris:97,ramirez:99,bramwell:01}. Recently, quantum variants of the spin ice have attracted great interest~\cite{molavian:07,onoda:09,onoda:10}, which include Tb$_2$Ti$_2$O$_7$~\cite{gardner:99,gardner:01,mirebeau:07}, Tb$_2$Sn$_2$O$_7$~\cite{mirebeau:07}, and Pr$_2$Sn$_2$O$_7$~\cite{matsuhira:02,zhou:08}, Pr$_2$Ir$_2$O$_7$~\cite{nakatsuji:06,machida:09}, Yb$_2$Ti$_2$O$_7$~\cite{cao:09}, and Er$_2$Ti$_2$O$_7$~\cite{cao:09}.

In the dipolar spin ice, large $\langle111\rangle$ Ising magnetic moments ($\sim10\mu_B$) of Dy$^{3+}$ and Ho$^{3+}$ ions, each of which points either inwards (``in'') to or outwards (``out'') from the center of the tetrahedron, interact with each other mainly through the magnetic dipole interaction~\cite{hertog:00}. This gives a ferromagnetic nearest-neighbor coupling, forming the ``2-in, 2-out'' ice rule~\cite{pauling} and leaving macroscopically degenerate ground states because of geometrical frustration. The spin dynamics is quite slow and the quantum effects are almost negligible.

Recently, quantum effects have been observed with inelastic neutron-scattering experiments on the spin-ice related compounds, Tb$_2$Ti$_2$O$_7$~\cite{gardner:99,gardner:01,mirebeau:07}, Tb$_2$Sn$_2$O$_7$~\cite{mirebeau:07}, and Pr$_2$Sn$_2$O$_7$~\cite{zhou:08}. 
Pr$_2TM_2$O$_7$ ($TM$ = Zr, Sn, and Ir) shows no magnetic dipole LRO but a (partial) spin freezing at $T_f\sim0.1$-$0.3$ K~\cite{matsuhira:02,zhou:08,nakatsuji:06,machida:09,matsuhira:09}. Pr$_2$Ir$_2$O$_7$ shows a metamagnetic transition at low temperatures only when the magnetic field is applied in the [111] direction~\cite{machida:09}, as in the spin ice. 
A chiral spin liquid~\cite{wen:89} has been detected through the anomalous Hall effect~\cite{ahe} at zero magnetic field without magnetic dipole LRO in Pr$_2$Ir$_2$O$_7$~\cite{machida:09}. Unlike the dipolar spin ice, the Curie-Weiss temperature $T_{CW}$ is antiferromagnetic for the zirconate~\cite{matsuhira:09} and iridate~\cite{nakatsuji:06}. The stannate shows a significant level of low-energy short-range spin dynamics in the energy range up to a few Kelvin~\cite{zhou:08}, which is absent in the classical spin ice.
Vital roles of the planar components have also been experimentally observed in Yb$_2$Ti$_2$O$_7$ and Er$_2$Ti$_2$O$_7$~\cite{cao:09}. 
Obviously, quantum fluctuations enrich the otherwise classical properties of the spin ice.

The mechanism in which the quantum fluctuations indicated by these observations appear depend on the rare-earth ions, including the difference in Kramers and non-Kramers ions. A relatively small localized magnetic moment reduces the magnetic dipole interaction, allowing for an important role of the superexchange interaction, as in the case of Pr$^{3+}$ ions~\cite{onoda:09,onoda:10} and also expected for Nd$^{3+}$, Sm$^{3+}$, and Yb$^{3+}$ ions because of their small moment amplitudes, $3.3\mu_B$, $0.7\mu_B$, and $4\mu_B$, respectively, for isolated cases. In the case of Tb$^{3+}$ ions, the presence of low-energy crystalline electric field levels also plays a role~\cite{molavian:07}. Remarkably, the large $LS$ coupling and the crystalline electric field for localized $4f$ electrons makes the superexchange interaction highly anisotropic~\cite{onoda:09,onoda:10}. In fact, the strength of the anisotropy depends on the properties of local magnetic doublets of rare-earth ions, producing the variety of experimentally observed low-temperature magnetic properties which are mentioned above.
Theoretically, it is expected that this anisotropic superexchange interaction drives {\it quantum phase transitions among the spin ice, quadrupolar states having nontrivial chirality correlations, and the quantum spin ice} for non-Kramers magentic doublets.

As for Kramers doublets, on the other hand, understandings of Yb$_2$Ti$_2$O$_7$ are rather controversial. Yb localized $4f$ magnetic moments are believed to almost lie perpendicular to the $\langle111\rangle$ directions, unlike Dy, Ho, and Pr cases. It has a ferromagnetic Curie-Weiss temperature $\sim0.65$~K and a first-order transition has been observed at 0.24 K with the specific heat measurement~\cite{blote:69}. It has been reported that a neutron-scattering experiment on the single crystal shows a LRO in the low-temperature phase~\cite{yasui:03}, while a combined study using neutron diffraction and M\"{o}ssbauer and $\mu$SR spectroscopies~\cite{hodges:02} and the polarized newtron-scattering~\cite{gardner:04} on the polycrystals have indicated that the magnetic dipole correlation remains dynamic. Recent neutron-scattering experiments have clarified a [111] rod scattering intensity in the paramagnetic phase~\cite{ross:09}, which has been phenomenologically analyzed by taking into account the anisotropic superexchange interaction~\cite{thompson:10}.

In this paper, we present a microscopic derivation of an effective quantum pseudospin-$1/2$ model, which takes the sum of the magnetic dipole interaction and the superexchange interaction, in terms of atomic Kramers doublets of Yb$^{3+}$ ions for the frustrated magnet Yb$_2TM_2$O$_7$ ($TM$ = Ti and Sn) on the pyrochlore lattice.

\section{Local Kramers doublets of Yb$^{3+}$ ions}

The ground state of an isolted Yb$^{3+}$ ion is characterized by the eightfold degenerate manifold $^2F_{\frac{7}{2}}$. In the pyrochlore oxide Yb$_2$Ti$_2$O$_7$, the $D_{3d}$ crystalline electric field partially lift the degeneracy, yielding the Kramers ground doublet
\begin{equation}
  |\sigma\rangle_D=-\alpha\sigma|J_z=\frac{7}{2}\sigma\rangle+\beta|J_z=\frac{1}{2}\sigma\rangle+\gamma\sigma|J_z=-\frac{5}{2}\sigma\rangle,
  \label{eq:doublet}
\end{equation}
with $\sigma=\pm$, real coefficients $\alpha\approx0.388$, $\beta\approx0.889$, and $\gamma\approx0.242$~\cite{hodges:01,note}, and the eigenstate $|J_z=M\rangle$ of the total angular momentum $J^z$, where the local coordinates $(\bm{x}_i,\bm{y}_i,\bm{z}_i)$ ($i=0,\cdots,3$) are site-dependent, as explicitly given in \ref{app:coordinates}. In particular, the quantization axis $\bm{z}_i$ is always taken to be parallel to the $\langle111\rangle$ direction that points from the Yb site to the center of one of the two tetrahedrons to which the Yb$^{3+}$ ion belongs. This ground doublet is energetically well separated from the first excitated doublet by  620~K~\cite{hodges:01}. 

The ground doublet has the following nonvanishing Ising and transverse components of the total angular momentum $\bm{J}$
\begin{eqnarray}
  _D\langle\sigma|\hat{J}^z|\sigma'\rangle_D&=&\frac{1}{2}J_\parallel\sigma^z_{\sigma,\sigma'},
  \label{eq:Jz}\\
  _D\langle\sigma|\hat{J}^{x,y} |\sigma'\rangle_D&=&\frac{1}{2}J_\bot\sigma^{x,y}_{\sigma,\sigma'},
  \label{eq:J+-}
\end{eqnarray}
where $J_\parallel=7\alpha^2+\beta^2-5\gamma^2$ and $J_\bot=2\sqrt{7}\alpha\gamma+4\beta^2$. On the other hand, it does not contribute to the atomic quadrupole moment, as is clear from
\begin{equation}
  _D\langle\sigma|\left\{\hat{J}^\pm,\hat{J}^z\right\}|\sigma'\rangle_D=
  _D\langle\sigma|\hat{J}^+ \hat{J}^+|\sigma'\rangle_D=_D\langle\sigma|\hat{J}^- \hat{J}^-|\sigma'\rangle_D=0.
  \label{eq:J+-Jz,J+-J+-}
\end{equation} 

Let us asign the Pauli matrices $\hat{\bm{\sigma}}_{\bm{r}}$ to a psedospin-$\frac{1}{2}$ that operates on the subspace for the the atomic doublet at the site $\bm{r}$. Then, the localized magnetic dipole moment is given by
\begin{equation}
  \hat{\bm{m}}_{\bm{r}}=g_J\mu_B\hat{\bm{J}}_{\bm{r}}=\frac{1}{2}\mu_B\left[g_\bot\left(\hat{\sigma}^x_{\bm{r}}\bm{x}_{\bm{r}}+\hat{\sigma}^y_{\bm{r}}\bm{y}_{\bm{r}}\right)+g_\parallel \hat{\sigma}^z_{\bm{r}}\bm{z}_{\bm{r}}\right],
\label{eq:m}
\end{equation}
where $g_\bot=g_JJ_\bot\approx4.18$ and $g_\parallel=g_JJ_\parallel\approx1.77$ are reasonably close to experimental values 4.27 and 1.79, respectively, for Yb$_2$Ti$_2$O$_7$~\cite{hodges:01}, with $g_J=8/7$ being the Land\'{e} factor.
This sharply contrasts to the non-Kramers magnetic doublet of Pr$^{3+}$ ions, where $\hat{\sigma}^z_{\bm{r}}$ only contributes to the magnetic dipole moment and $\hat{\sigma}^\pm_{\bm{r}}$ to the magnetic quadrupole moment~\cite{onoda:09,onoda:10}.

\section{Magnetic dipole interaction}

The magnetic dipole interaction takes the form
\begin{equation}
  \hat{H}_{\mathrm{D}}=\frac{\mu_0}{4\pi}\sum_{\langle\bm{r},\bm{r}'\rangle}\left[\frac{\hat{\bm{m}}_{\bm{r}}\cdot\hat{\bm{m}}_{\bm{r}'}}{({\mit\Delta}r)^3}-3\frac{(\hat{\bm{m}}_{\bm{r}}\cdot{\mit\Delta}\bm{r})({\mit\Delta}\bm{r}\cdot\hat{\bm{m}}_{\bm{r}'})}{({\mit\Delta}r)^5}\right],
  \label{eq:H_D}
\end{equation}
with ${\mit\Delta}\bm{r}=\bm{r}-\bm{r}'$ and the summation $\sum_{\langle\bm{r},\bm{r}'\rangle}$ over all the pairs of atomic sites, where the localized magnetic moment is given by Eq.~\eqref{eq:m}.

\section{Superexchange interaction}

\begin{figure}[tb]
\begin{center}
\includegraphics[width=36pc]{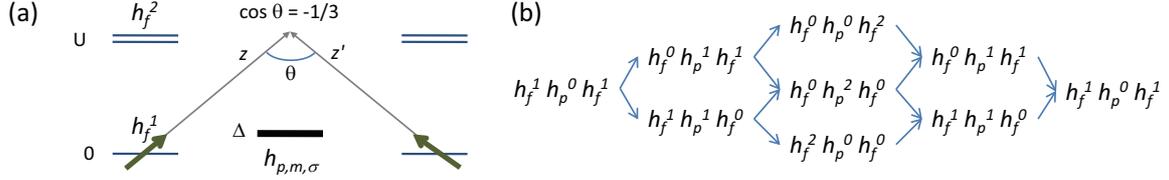}\hspace{2pc}%
\end{center}
\caption{\label{fig:perturbation}(a) Local level scheme of O $2p$ holes and Yb $4f$ holes. (b) Vitual hopping processes of holes contributing to the superexchange interaction.  $n$ ($n'$) and $\ell$ in the state $h_f^nh_p^\ell h_f^{n'}$ represent the number of $f$ holes at the Yb site $\bm{r}$ ($\bm{r}'$) and that of $p$ holes at the O site.}
\end{figure}

To derive an anisotropic superexchange interaction between Yb magnetic moments,
it is convenient to take the hole picture, where the local ground state is described by a single $4f$ hole at each Yb site and no hole at O stites. The local Coulomb repulsion among Yb $4f$ holes and the energy level $\Delta$ of the O $2p$ holes compared with that of a Yb $4f$ hole provide the largest energy scales of the problem, as schematically shown in Fig.~\ref{fig:perturbation} (a). In fact, the local ground state is modified by the nonlocal electron transfer. The amplitude is described by two Slater-Koster parameters $V_{pf\pi}$ and $V_{pf\sigma}$, which represent the transfer integrals between $p_x$/$p_y$ and $f_{x(5z^2-r^2)}$/$f_{y(5z^2-r^2)}$ orbitals and between $p_z$ and $f_{(5z^2-3r^2)z}$ orbitals, respectively~\cite{sharma:79}.
Then, we perform the fourth-order strong-coupling perturbation expansion of the $f$-$p$ electron transfer, by fully taking into account the virtual electron transfer processes among O $2p$ orbitals located at the center of the tetrahedron and $4f$ orbitals located at adjacent Yb sites, which are shown in Fig.~\ref{fig:perturbation} (b), in the local coordinate frames. The superexchange Hamiltonian is then obtained as
\begin{eqnarray}
  \hat{H}_{ff}=\frac{2}{\Delta^2}\sum_{\langle\bm{r},\bm{r}'\rangle}^{\mathrm{n.n.}}\sum_{m_1,m_2}\sum_{m_1',m_2'}\sum_{\sigma_1,\sigma_2}\sum_{\sigma_1',\sigma_2'}
    V_{m_1}V_{m_1'}V_{m_2}V_{m_2'}
    \hat{f}^\dagger{}_{\bm{r},m_1,\sigma_1}\hat{f}_{\bm{r},m_2,\sigma_2}
    \hat{f}^\dagger{}_{\bm{r}',m_1',\sigma_1'}\hat{f}_{\bm{r}',m_2',\sigma_2'}
  \nonumber\\
  \times(\frac{1}{\Delta}+\frac{1}{U})\left(R^\dagger_{\bm{r}}R_{\bm{r}'}\right)_{m_1,m_2';\sigma_1,\sigma_2'}\left(R^\dagger_{\bm{r}'}R_{\bm{r}}\right)_{m_1',m_2;\sigma_1',\sigma_2}.
  \label{eq:H_sex}
\end{eqnarray}
with $V_{\pm1}=V_{pf\pi}$ and $V_0=V_{pf\sigma}$ as well as the rotation operator $R_{\bm{r}}$ of the angular momentum from the global coordinates to the local ones at the Yb site $\bm{r}$, which is given in \ref{app:coordinates}. Here, $\hat{f}^\dagger{}_{\bm{r},m,\sigma}$ and $\hat{f}_{\bm{r},m,\sigma}$ represent the creation (annihilation) and annihilation (creation) operators of $4f$ electron (hole) with the orbital $m$ and the spin $\sigma$ at the Yb site $\bm{r}$. $R_{\bm{r}}$ is the rotation matrix of the total angular momentum from the glocal coordinates to the local, as given in \ref{app:coordinates}. Projecting Eq.~\eqref{eq:H_sex} onto the space of the doublets, we obtain
\begin{eqnarray}
  \hat{H}_{ex}&=&-J_{n.n.}\sum^{n.n.}_{\langle\bm{r},\bm{r}'\rangle}\left[
    g^\parallel\hat{\sigma}^z_{\bm{r}}\hat{\sigma}^z_{\bm{r}'}
    +g^\bot\left(\hat{\sigma}^x_{\bm{r}}\hat{\sigma}^x_{\bm{r}'}+\hat{\sigma}^y_{\bm{r}}\hat{\sigma}^y_{\bm{r}'}\right)
    \right.\nonumber\\
    &&\left.\ \ \ \ \ \ \ \ \ \ \ \ \ \ 
    +g^q\left(\left(\hat{\vec{\sigma}}_{\bm{r}}\cdot\vec{n}_{\bm{r}}\right)\left(\hat{\vec{\sigma}}_{\bm{r}'}\cdot\vec{n}_{\bm{r}'}\right)
    -\left(\hat{\vec{\sigma}}_{\bm{r}}\cdot\vec{n}'_{\bm{r},\bm{r}'}\right)\left(\hat{\vec{\sigma}}_{\bm{r}'}\cdot\vec{n}'_{,\bm{r},\bm{r}'}\right)\right)
    \right.\nonumber\\
    &&\left.\ \ \ \ \ \ \ \ \ \ \ \ \ \ 
    +g^K\left(\hat{\sigma}^z_{\bm{r}}\left(\hat{\vec{\sigma}}_{\bm{r}'}\cdot\vec{n}_{\bm{r},\bm{r}'}\right)+\left(\hat{\vec{\sigma}}_{\bm{r}}\cdot\vec{n}_{\bm{r},\bm{r}'}\right)\hat{\sigma}^z_{\bm{r}'}\right)
    \right],
  \label{eq:H_ex}
\end{eqnarray}
with $J_{n.n.}=\frac{(2\beta V_{pf\sigma})^4}{3^3(7\Delta)^2}\left(\frac{1}{U}+\frac{1}{\Delta}\right)$ and the dimensionless coupling constants as functions of $x=V_{pf\pi}/V_{pf\sigma}$,
\begin{eqnarray}
  g^\parallel&=&1-8\sqrt{6}x-\frac{9}{2}x^2-3\sqrt{6}x^3+\frac{63}{16}x^4,
  \label{eq:g^parallel}\\
  g^\bot&=&1+4\sqrt{6}x+\frac{45}{2}x^2-3\sqrt{6}x^3+\frac{9}{16}x^4,
  \label{eq:g^bot}\\
  g^q&=&-2\left(1-2\sqrt{6}x+9x^2-3\sqrt{6}x^3+\frac{9}{4}x^4\right),
  \label{eq:g^q}\\
  g^K&=&2\sqrt{2}\left(1+\sqrt{6}x-\frac{45}{4}x^2+\frac{15}{4}\sqrt{6}x^3-\frac{9}{8}x^4\right).
  \label{eq:g^K}
\end{eqnarray}
Here, we have introduced $\hat{\vec{\sigma}}_{\bm{r}}=(\hat{\sigma}^x_{\bm{r}},\hat{\sigma}^y_{\bm{r}})$ as well as two unit vectors $\vec{n}_{\bm{r},\bm{r}'}=(\cos\phi_{\bm{r},\bm{r}'},-\sin\phi_{\bm{r},\bm{r}'})$ and $\vec{n}_{\bm{r},\bm{r}'}'=(\sin\phi_{\bm{r},\bm{r}'},\cos\phi_{\bm{r},\bm{r}'})$. In the choice of the local coordinate frames given in \ref{app:coordinates}, the phase takes $\phi_{\bm{r},\bm{r}'}=-2\pi/3$, $2\pi/3$, and 0, when the pair of Yb sites $\bm{r}$ and $\bm{r}'$ corresponds to that of $\bm{R}+\bm{a}_i$ and $\bm{R}+\bm{a}_j$ with $(i,j)=\{(0,1),(2,3)\}$, $\{(0,2),(1,3)\}$, and $\{(0,3),(1,2)\}$, respectively, where $\bm{a}_0=-\frac{a}{8}(1,1,1)$, $\bm{a}_1=\frac{a}{8}(-1,1,1)$, $\bm{a}_2=\frac{a}{8}(1,-1,1)$, and $\bm{a}_3=\frac{a}{8}(1,1,-1)$ with the lattice constant $a$ and an fcc lattice vector $\bm{R}$.

Previously, this form of the nearest-neighbor Hamiltonian with and without the $J^K_{n.n.}$ term is found from the symmetry analysis for the Kramers doublets~\cite{onoda:10} and microscopically for the non-Kramers doublets~\cite{onoda:09,onoda:10}, respectively. The present microscopic calculation of the exchange coupling constants for the Kramers doublets of Yb$^{3+}$ ions remarkably reveals that the symmetric exchange part is ferroic in terms of pseudospins, i.e., ``in'' or ``out'', and antiferromagnetic in terms of real magnetic moments, unlike the case of Pr moments~\cite{onoda:09,onoda:10}, and that the four coupling constants are of the same order in the magnitude with their ratios controlled by a single parameter $V_{pf\pi}/V_{pf\sigma}$.

\begin{figure}[tb]
\begin{center}
\includegraphics[width=24pc]{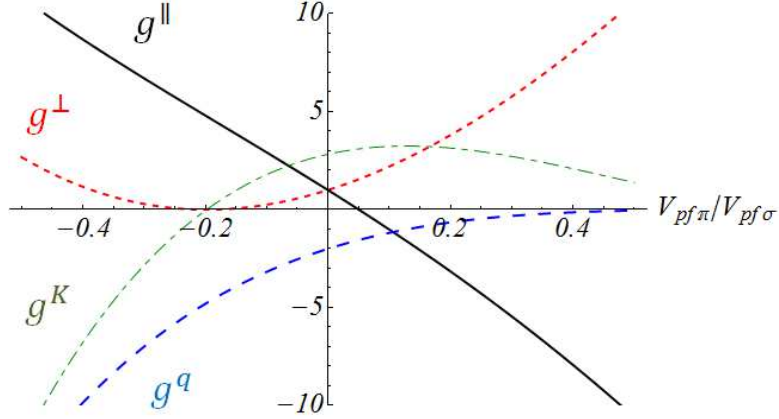}\hspace{2pc}
\end{center}
\caption{\label{fig:couplings}Dimensionless coupling constants as functions of $V_{pf\pi}/V_{pf\sigma}$.}
\end{figure}

\section{Summary}
We have obtained a highly anisotropic effective pseudospin-$1/2$ Hamiltonian for interacting Yb Kramers doublets of Yb$_2TM_2$O$_7$, which includes four adjustable parameters; $\alpha$ and $\gamma$ describing crystal-field doublets, $J_{n.n.}$ for the magnitude of whole superexchange terms, and $V_{pf\pi}/V_{pf\sigma}$ specifying the ratios of aniotropic exchange couplings. Theoretical analyses of the model in comparison with experiments will be published elsewhere.

\ack
The author thanks M. J. P. Gingras for a critical reading of the manuscript. The work was partly supported by Grants-in-aid for Scientific Research under Grant No. 19052006 from the Ministry of Education, Culture, Sports, Science, and Technology of Japan and No. 21740275 from Japan Society for the Promotion of Science.

\appendix
\section{Local coordinate frames}
\label{app:coordinates}

The local coordinate frames are chosen as
\begin{eqnarray}
  \bm{x}_0=-\frac{1}{\sqrt{6}}\left(1,1,-2\right),
  \ \ \
  \bm{y}_0=-\frac{1}{\sqrt{2}}\left(-1,1,0\right),
  \ \ \ 
  \bm{z}_0=\frac{1}{\sqrt{3}}(1,1,1),
  \label{eq:xyz0}\\
  \bm{x}_1=-\frac{1}{\sqrt{6}}\left(1,-1,2\right),
  \ \ \ 
  \bm{y}_1=-\frac{1}{\sqrt{2}}\left(-1,-1,0\right),
  \ \ \
  \bm{z}_1=\frac{1}{\sqrt{3}}(1,-1,-1),
  \label{eq:xyz1}\\
  \bm{x}_2=-\frac{1}{\sqrt{6}}\left(-1,1,2\right),
  \ \ \ 
  \bm{y}_2=-\frac{1}{\sqrt{2}}\left(1,1,0\right),
  \ \ \ 
  \bm{z}_2=\frac{1}{\sqrt{3}}(-1,1,-1),
  \label{eq:xyz2}\\
  \bm{x}_3=-\frac{1}{\sqrt{6}}\left(-1,-1,-2\right),
  \ \ \ 
  \bm{y}_3=-\frac{1}{\sqrt{2}}\left(1,-1,0\right),
  \ \ \ 
  \bm{z}_3=\frac{1}{\sqrt{3}}(-1,-1,1),
  \label{eq:xyz3}
\end{eqnarray}
for the Yb sites at $\bm{R}+\bm{a}_i$ ($i=0,\cdots,3$).
They are invariant under the twofold rotations about $x$, $y$, and $z$ axes that include the center of the tetrahedron. In particular, all the local $z$ axes given above point outwards from the center of the tetrahedron. 
The coordinate frame for the spins is always attached to that for the orbital space in each case. The rotation matrix of the total angular momentum $\bm{j}=\bm{l}+\bm{s}$ with the orbital $\bm{l}$ and the spin $\bm{s}$ of a single electron takes the form
\begin{equation}
  \hat{R}_{\bm{r}}=\exp\left[-i\varphi_i\hat{j}^z\right]\exp\left[-i\vartheta_i\hat{j}^y\right]\exp\left[-i\pi\hat{j}^z\right].
  \label{eq:R}
\end{equation}
with
$\varphi_0=\pi/4$,
$\vartheta_0=\arccos\left(1/\sqrt{3}\right)$,
$\varphi_1=3\pi/4$,
$\vartheta_1=-\pi+\arccos\left(1/\sqrt{3}\right)$,
$\varphi_2=-\pi/4$,
$\vartheta_2=-\pi+\arccos\left(1/\sqrt{3}\right)$,
$\varphi_3=-3\pi/4$, and
$\vartheta_3=\arccos\left(1/\sqrt{3}\right)$.

\section{Single-hole $f$-electron states}
\label{app:hole}

The Kramers doublet given in Eq.~\eqref{eq:doublet} can be expressed in terms of single $f$-hole states $|l^z,s^z\rangle_h$ with the orbital $l^z$ and the spin $s^z$ using
\begin{eqnarray}
  |J^z=\frac{7}{2}\sigma\rangle
  &=&
  |-3\sigma,-\frac{\sigma}{2}\rangle_h,
  \nonumber\\
  |J^z=\frac{5}{2}\sigma\rangle
  &=&
  \frac{1}{\sqrt{7}}|-3\sigma,\frac{\sigma}{2}\rangle_h
  +\sqrt{\frac{6}{7}}|-2\sigma,-\frac{\sigma}{2}\rangle_h,
  \nonumber\\
%  |J^z=\frac{3}{2}\sigma\rangle
%  &=&
%  \sqrt{\frac{2}{7}}|-2\sigma,\frac{\sigma}{2}\rangle_h
%  +\sqrt{\frac{5}{7}}|-\sigma,-\frac{\sigma}{2}\rangle_h
%  \nonumber\\
  |J^z=\frac{\sigma}{2}\rangle
  &=&
  \sqrt{\frac{3}{7}}|-\sigma,\frac{\sigma}{2}\rangle_h
  +\frac{2}{\sqrt{7}}|0,-\frac{\sigma}{2}\rangle_h.
\end{eqnarray}

\section*{References}
%%%%%%%%%%%%%%%%%%%%%%%%%%%%%%%%%%%%%%%%%%%


\begin{thebibliography}{99}
\bibitem{anderson:56}
  Anderson P W 1956 {\it Phys. Rev.} \textbf{102} 1008

\bibitem{anderson}
  Anderson P W 1973 {\it Mater. Res. Bull.} \textbf{8} 153

\bibitem{wen:89}
  Wen X G, Wilczek F and Zee A 1989 {\it Phys. Rev.} B \textbf{39} 11413

% Heisenberg AF
\bibitem{moessner:98}
  R. Moessner and J. T. Chalker, {\it Phys. Rev. Lett.} \textbf{80}, 2929 (1998).

\bibitem{isakov:04}
  Isakov S V, Gregor K, Moessner R and Sondhi S L 2004 {\it Phys. Rev. Lett.} \textbf{93} 167204

\bibitem{gardner:10}
  Gardner J S, Gingras M J P and Greedan J E, {\it Rev. Mod. Phys.} \textbf{82} 53

%Spin ice
\bibitem{harris:97}
  Harris M J, Bramwell S T, McMorrow D F, Zeiske T and Godfrey K W 1997 {\it Phys. Rev. Lett.} \textbf{79} 2554
\bibitem{ramirez:99}
  Ramirez A P {\it et al.} 1999 {\it Nature (London)} \textbf{399} 333
\bibitem{bramwell:01}
  Bramwell S T and Gingras M J P 2001 {\it Science} \textbf{294} 1495


%Quantum spin ice
\bibitem{molavian:07}
  Molavian H R, Gingras M J P and Canals B 2007 {\it  Phys. Rev. Lett.} \textbf{98} 157204

\bibitem{onoda:09}
  Onoda S and Tanaka Y 2010 {\it Phys. Rev. Lett.} \textbf{105} 047201

\bibitem{onoda:10}
  Onoda S and Tanaka Y 2010 {\it Preprint} arXiv:1011.4981

%Tb2Ti2O7
\bibitem{gardner:99}
  Gardner J S {\it et al.} 1999 {\it Phys. Rev. Lett.} \textbf{82} 1012
\bibitem{gardner:01}
  Gardner J S, Gaulin B D, Berlinsky A J, Waldron P, Dunsiger S R, Raju N P and Greedan J E 2001 {\it Phys. Rev.} B \textbf{64} 224416
\bibitem{mirebeau:07}
  Mirebeau I, Bonville P and Hennion M 2007 {\it Phys. Rev.} B \textbf{76} 184436

%Pr pyrochlore Pr2Sn2O7 Pr2Ir2O7
\bibitem{matsuhira:02}
  Matsuhira K {\it et al.} 2002 {\it J. Phys. Soc. Jpn.} \textbf{71} 1576

\bibitem{zhou:08}
  Zhou H D, Wiebe C R, Janik J A, Balicas L, Yo Y J, Qiu Y, Copley J R D and Gardner J S 2008 {\it Phys. Rev. Lett.} \textbf{101} 227204

\bibitem{nakatsuji:06}
  Nakatsuji S {\it et al.} 2006
  {\it Phys. Rev. Lett.} \textbf{96} 087204

\bibitem{machida:09}
  Machida Y, Nakatsuji S, Onoda S, Tayama T and Sakakibara T 2010 {\it Nature (London)} \textbf{463} 210

\bibitem{cao:09}
  Cao H, Gukasov A, Mirebeau I, Bonville P, Decorse C and Dhalenne G 2009 {\it Phys. Rev. Lett.} \textbf{103} 056402

\bibitem{hertog:00}
  den Hertog B C and Gingras M J P 2000 {\it Phys. Rev. Lett.} \textbf{84} 3430

\bibitem{pauling}
  Pauling L 1938 {\it The Nature of the Chemical Bonds} (Ithaca: Cornel University Press)

%Pr2Zr2O7
\bibitem{matsuhira:09}
  Matsuhira K {\it et al.} 2009 {\it  J. Phys.: Conf. Series} \textbf{145} 012031

\bibitem{ahe}
  Nagaosa N, Sinova J, Onoda S, MacDonald A H and Ong  N P 2010 {\it Rev. Mod. Phys.} \textbf{82} 1539

\bibitem{blote:69}
  Bl\"{o}te W J and Wielinga R F and Huiskamp W J 1969 {\it Physica (Amsterdam)} \textbf{43} 549

\bibitem{yasui:03}
  Yasui Y {\it et al.} 2003 {\it J. Phys. Soc. Jpn.} \textbf{72} 3014

\bibitem{hodges:02}
  Hodges J A {\it et al.} 2002 {\it Phys. Rev. Lett.} \textbf{88} 077204

\bibitem{gardner:04}
  Gardner J, Ehlers G, Rosov N, Erwin R W and Pertovic C 2004 {\it Phys. Rev.} B \textbf{70} 180404(R)

\bibitem{ross:09}
  Ross K A, Ruff J P C, Adams C P, Gardner J S, Dabkowska H A, Qiu Y, Copley J R D and Gaulin B D 2009 {\it Phys. Rev. Lett.} \textbf{103} 227202.

\bibitem{thompson:10}
  Thompson J D, McClarty P A, Ronnow H M, Regnault L P, Sorge A and M J P Gingras 2010 {\it Preprint} arXiv:1010.5476.

\bibitem{hodges:01}
  Hodges J A, Bonville P, Forget A, Rams M, Kr\'{o}las K and Dhalenne G 2001 {\it J. Phys.: Condens. Matter} \textbf{13} 9301

\bibitem{note}
  Similar CEF parameters have been obtained in Ref.~\cite{cao:09}.

%Slater-Koster
\bibitem{sharma:79}
  Sharma R R, Phys. Rev. B \textbf{19}, 2813 (1979).

\end{thebibliography}
\end{document}